\documentclass[iop,numberedappendix]{emulateapj}

\usepackage{url}
\usepackage{enumerate}
\usepackage{natbib}

\newcommand{\sige}{$\sigma_{\rm e}$}

\newcommand{\her}{\emph{Herschel}}

\newcommand{\nh}{$N_{\rm H}$}
\newcommand{\av}{$A_{V}$}

\newcommand{\IRAS}{\textit{IRAS}}
\newcommand{\Planck}{\textit{Planck}}

\newcommand{\Ehk}{$E(H - K_{\rm s})$}

\newcommand{\Ejh}{$E(J-H)$}
\newcommand{\Ejk}{$E(J-K_{\rm s})$}

\newcommand{\lsol}{$L_{\odot}$}

\newcommand{\msol}{$M_{\odot}$}

\slugcomment{To appear in the Astrophysical Journal}
\begin{document}

\shorttitle{Dust opacity in Orion}

\shortauthors{Roy et al.}


\title{ Changes of Dust Opacity with Density in the Orion A Molecular Cloud}


\author{Arabindo~Roy,\altaffilmark{1}
        Peter~G.~Martin,\altaffilmark{1}
        Danae~Polychroni,\altaffilmark{2}
        Sylvain~Bontemps,\altaffilmark{3}
        Alain~Abergel,\altaffilmark{4}
        Philippe~Andr\'e,\altaffilmark{5}
        Doris~Arzoumanian,\altaffilmark{5}
        James~Di~Francesco,\altaffilmark{6}
        Tracey~Hill,\altaffilmark{5}
        Vera~Konyves,\altaffilmark{4,5}
        Quang~Nguyen-Luong,\altaffilmark{1}   
        Stefano~Pezzuto,\altaffilmark{7}
        Nicola~Schneider,\altaffilmark{3,5}
        Leonardo~Testi,\altaffilmark{8}  
        and Glenn~ White\altaffilmark{9,10}
}

\altaffiltext{1}{Canadian Institute for Theoretical Astrophysics, University of Toronto, 60 St. George Street, Toronto, ON M5S~3H8, Canada}
\altaffiltext{2}{INAF-IFSI, via Fosso del Cavaliere 100, 00133 Roma, Italy}
\altaffiltext{3}{Universit\'e de Bordeaux, LAB, UMR5804, 33270 Floirac, France}
\altaffiltext{4}{IAS, CNRS (UMR 8617), Universit\'e Paris-Sud 11, B\^atiment 121, 91400 Orsay, France }`
\altaffiltext{5}{Laboratoire AIM, C.E.A. Saclay, 90091 Gif-sur-Yvette, France}
\altaffiltext{6}{National Research Council of Canada, 5071 West
Saanich Road, Victoria, BC V9E 2E7, Canada}
\altaffiltext{7}{Istituto di Astrofisica e Planetologia Spaziali – IAPS, Istituto Nazionale di Astrofisica – INAF, via Fosso del Cavaliere 100,
00133 Roma, Italy}
\altaffiltext{8}{European Southern Observatory, Karl Schwarzschild Strasse 2, D-85748 Garching, Germany}
\altaffiltext{9}{Department of Physics and Astronomy, The Open University, Walton Hall, Milton Keynes, MK7 6AA, UK}
\altaffiltext{10}{RAL Space, STFC Rutherford Appleton Laboratory, Chilton, Didcot, Oxfordshire, OX11 0QX, UK}



\begin{abstract}
  
We have studied the opacity of dust grains at submillimeter
wavelengths by estimating the optical depth from imaging at 160,
250, 350, and 500~\micron\ from the \her\ Gould Belt 
Survey and comparing this to a column density obtained from the 2MASS-derived
color excess \Ejk.  Our main goal was to investigate the spatial
variations of the opacity due to `big' grains over a variety of
environmental conditions and thereby quantify how emission properties
of the dust change with column (and volume) density.  The central and
southern areas of the Orion~A molecular cloud examined here, with
$N_{\rm H}$ ranging from 1.5 $\times 10^{21}$ cm$^{-2}$ to 
$50 \times 10^{21}$ cm$^{-2}$, are well suited to this approach.
We fit the multi-frequency \her\ spectral energy distributions (SEDs)
of each pixel with a modified blackbody to obtain the temperature, $T$,
and optical depth, $\tau_{1200}$, at a fiducial frequency of 1200~GHz
(250~\micron).  Using a calibration of $N_{\rm H}/E(J-K_{s})$
for the interstellar medium (ISM) we obtained the opacity (dust
emission cross-section per H nucleon), \sige(1200), for every pixel.
From a value $\sim1 \times 10^{-25}$~cm$^2$~H$^{-1}$ at the lowest
column densities that is typical of the high latitude diffuse ISM,
\sige(1200) increases as $N_{\rm H}^{0.28}$ over the range studied.
This is suggestive of grain evolution.
Integrating the SEDs over frequency, we also calculated the specific
power $P$ (emission power per H) for the big grains.  In low column
density regions where dust clouds are optically thin to the
interstellar radiation field (ISRF), $P$ is typically $3.7\times
10^{-31}$~W~H$^{-1}$, again close to that in the high latitude diffuse
ISM.  However, we find evidence for a decrease of $P$ in high column
density regions, which would be a natural outcome of attenuation of
the ISRF that heats the grains, and for localized increases for dust
illuminated by nearby stars or embedded protostars.

\end{abstract}



\section{Introduction}

Thermal dust emission is optically thin at submillimeter
wavelengths. As such, it provides a useful probe of the 
interstellar medium (ISM) and the embedded filamentary 
and clumpy structures within it which relate
to the early stages of star formation.
These structures are being revealed in exquisite detail by 
the \emph{Herschel} Gould Belt Survey (HGBS, \citealp{andre2010}), 
the \emph{Herschel} imaging survey of OB Young Stellar objects (HOBYS,
\citealp{motte2010}), and
the \emph{Herschel} infrared Galactic Plane Survey (Hi-GAL,
\citealp{molinari2010}).
The best spatial resolution is obtained by the HGBS given the relative
proximity of the target molecular clouds.

For quantitative measurements of the total column density, \nh, and
for assessment of the mass and gravitational (in)stability of any
structures in the molecular cloud, the dust opacity is required.
Evidence for significant environmental changes in opacity, ranging
over an order of magnitude, has already been presented; see, e.g.,
\cite{juvela2011,abergelDd2011,abergelTd2011,Martin2012}.  Using new
HGBS data, our goal here is to investigate whether or not there is any
systematic dependence of opacity on column (and volume) density.

Theoretically, dust grains are expected to evolve in a sufficiently
dense medium, by aggregation or growth of ice mantles.  Certainly
this is the conclusion of a number of numerical simulations of dust
evolution in dense ISM environments (e.g., \citealp{ossenkopf1994,ormel2011}).
 Given the complexity of dust in the ISM,  however, a completely 
\emph{ab initio} prediction for a particular environment is challenging, 
and so it is useful to seek observational constraints on the onset 
and magnitude of any environmental changes.

Volume density, not column density \emph{per se}, ought to be the
determinant of dust evolution.  For this study we have selected the
southern part of the Orion~A molecular cloud imaged by the HGBS
(Polychroni et al. 2012  in preparation).
Its high column density, despite its
relatively high Galactic latitude, $b$ $\sim$ $-$19\degr, suggests
that the emission comes largely from a single cloud, in which case the
high column density can be reasonably attributed to high volume density.
Also, at least to a first approximation, the cloud should be bathed in
a relatively uniform interstellar radiation field (ISRF).  For
this sort of analysis, such a field is therefore much more favorable 
than one in the Galactic plane where a wide range of conditions would be
superimposed along the line of sight.

The larger (`big') dust grains, which account for most of the dust mass,
are in thermal equilibrium with the ambient radiation field
\citep{compiegne2011}.  The intensity of the thermal dust emission,
when optically thin, is given by
\begin{equation}
I_\nu = \tau_\nu  B_{\nu}(T) \equiv \sigma_{\rm e}(\nu) N_{\rm H} B_{\nu}(T),
\label{emissmass}
\end{equation}
where $\tau_\nu$ is the dust optical depth of the column of material,
$B_{\nu}(T)$ is the Planck function for dust temperature $T$,
$N_{\rm{H}}$ is the total hydrogen column density (H in any form), and
$\sigma_{\rm e}(\nu) = \tau_\nu/N_{\rm H}$ is the opacity of the
interstellar material (the emission cross-section per H
nucleon).\footnote{Note the correspondence $\sigma_{\rm e} (\nu)
  \equiv \mu m_{\rm H} r \kappa_\nu$ where $\kappa_\nu$ is the mass
  absorption (or emission) cross-section per gram of dust, $r$ is
  dust-to-gas mass ratio, and $\mu$ is the mean weight per H (1.4).
  Only the product $r \kappa_\nu$ is needed; this quantity is often
  also called the opacity, now in the alternate units cm$^2$g$^{-1}$.}

Clearly, quantifying the optical depth and the opacity requires
knowledge of $T$.  Determination of $T$ is made possible by the multi-frequency
coverage now available with the \emph{Herschel Space Observatory}
\citep{pilbratt2010}, through fitting the spectral energy distribution
(SED) of $I_\nu$.
The other requirement is an independent measure of \nh.  Here we have
used the near-infrared color excess \Ejk\ as a proxy
(see Appendix~\ref{sec:excess}),
derived from the Two Micron All Sky Survey (2MASS\footnote{The Two
Micron All Sky Survey (2MASS) is a joint project of the University of
Massachusetts and the Infrared Processing and Analysis
Center/California Institute of Technology, funded by the National
Aeronautics and Space Administration and the National Science
Foundation.})  data.

Another quantity that we examine (Section~\ref{sec:power}) is the
`specific power,' i.e., the total power per H emitted by dust grains in
thermal equilibrium:
\begin{equation}
P =  \int  4 \pi \sigma_{\rm e}(\nu) B_\nu(T) d\nu.
\label{power}
\end{equation}
Since in equilibrium $P$ is equal to the total energy absorbed per
H, $P$ is affected by the intensity of the ISRF, which can be higher
than average near a strong source of radiation, or lower than average
because of attenuation in a region of high column density (without
internal sources of illumination).  Note how, for a given absorbed
$P$, the resulting equilibrium temperature is inversely related to how
efficiently grains can emit.

This paper is organized as follows.
In Section~\ref{sec:obs}, we briefly describe \emph{Herschel} imaging
of the Orion~A molecular cloud using the  PACS \citep{poglitsch2010} 
and SPIRE \citep{griffin2010} cameras.
Maps of the SED fitting parameters $\tau$ and $T$ are derived in
Section~\ref{sec:taut}.  We discuss the SED fitting, validate the fits
through prediction of the 100~\micron\ emission for comparison with
observations by \IRAS\ (Appendix~\ref{appen:iris}), and assess the
uncertainty in $I_\nu$, and estimate the uncertainties of the derived
parameters through Monte-Carlo simulation (Appendix~\ref{sec:errors}).
In Section~\ref{sec:results}, we compare $\tau$ with \nh\ to find an
estimate of the opacity (Equation~(\ref{emissmass})).  It appears that
the opacity grows systematically with \nh, evidence for grain evolution.
We examine the dependence of quantities related to $P$ in
Section~\ref{sec:power}, specifically  the anti-correlation between 
$T$ and $\tau$ and the spatial and \nh\ dependence of $P$.
This analysis provides insight into the various interrelationships between
$T$, \sige, and $P$ discussed in Section~\ref{sec:inter}.
We conclude with a summary in Section~\ref{sec:conclusion}.


\section{\emph{Herschel} Observations}\label{sec:obs}

As part of the HGBS, three fields in the Orion~A molecular cloud were
mapped at a scanning speed of 60\arcsec\ s$^{-1}$ in parallel mode,
acquiring data simultaneously in five bands using the PACS
\citep{poglitsch2010} and SPIRE \citep{griffin2010} arrays.
Images were produced by the ROMAGAL map-maker \citep{traficante2011}
and first results on the filamentary and core substructures in the two
fields studied here, called Orion~A-C1 and Orion~A-S1, are presented
by Polychroni et al.\ (2012).
The images have angular resolutions of 9\farcs6, 13\farcs5, 18\farcs0,
24\farcs0, and 37\farcs0 at 70, 160, 250, 350, and 500 ~\micron,
respectively.  In the following analysis we did not use the
70~\micron\ image.
Zero-point offsets added to these images, obtained by correlating with
\Planck\ and \IRAS\ images \citep{bernard2010}, were 
8, 20, 10, and 4~MJy~sr$^{-1}$ at 160, 250, 350, and 500~\micron,
respectively for the Orion~A-C1 images and similarly
3, 14, 18, and 8~MJy~sr$^{-1}$ for the Orion~A-S1 images.
Uncertainties in the zero-point offsets have the most effect on the
SEDs of pixels with lower brightnesses, and the offsets were refined
as discussed in Appendix~\ref{sec:errors}.
Prior to fitting SEDs for each pixel, we convolved and then regridded
the individual images to correspond to the lowest resolution
(37\farcs0) on a common grid with 11\farcs5 pixels.
Power spectra of these images follow a `cirrus-like' power-law
relation that decays with spatial frequency until at the highest
frequencies it merges with a flat level corresponding to the noise in
the map \citep{roy2010,martin2010}, assessed at  
0.92, 0.58, 0.31, and 0.38~MJy~sr$^{-1}$ for 160, 250, 350, and
500~\micron, respectively.


\section{Maps of $\tau$ and $T$}\label{sec:taut}



\begin{figure}
\includegraphics[width=\linewidth]{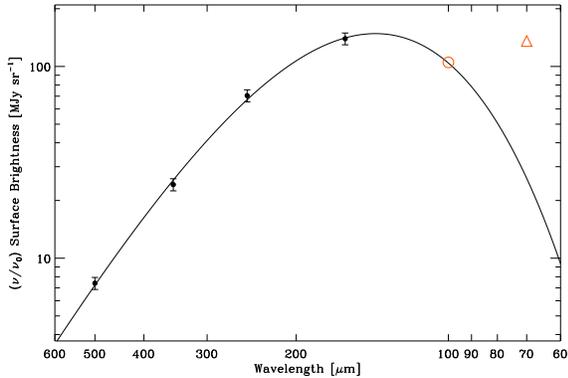}
\caption{Representative SED in the Orion~A map.  The derived
  temperature is 17.9~K and the optical depth at 250~\micron\ is
  6.3$\times10^{-4}$.  Only the four lowest frequency \her\ passbands
  were used in fitting the modified blackbody SED.  This model predicts the
  100~\micron\ \IRAS\ brightness quite well (open circle, see
  Appendix~\ref{appen:iris}) but underpredicts the 70~\micron\ emission
  (triangle) where non-equilibrium emission from VSGs becomes
  important.}
\label{fig:sed}
\end{figure}

We parametrized the spectral dependence of the opacity (or $\tau$) as
$\sigma_{\rm e}(\nu)/\sigma_{\rm e} (\nu_0) = (\nu/\nu_0)^{\beta}$,
with a fiducial frequency $\nu_0 = 1200$~GHz (250~\micron).  Although 
HGBS  consortium assumed an emissivity index, $\beta$, of 2.0, however,
in this paper we adopted a fixed $\beta =1.8$ to facilitate comparison 
with earlier analyses \citep{abergelDd2011,Martin2012}.
Treating $\beta$ as a fitting parameter does not change our results
(Section~\ref{sec:tautnh}).
The SEDs are thus fit with two parameters, $\tau_{1200}$ and $T$.
In our SED fits, we have
not used 70~\micron\ PACS data because of probable contamination due to
non-equilibrium emission by smaller dust grains (Very Small Grains, or
VSGs) which broadens the apparent SED toward wavelengths shortward of
the spectral peak.  The SED of cold dust emission for $\beta =1.8$ and
a temperature of 15~K, typical of this cloud, peaks at 200~\micron,
and so the remaining four submillimeter \emph{Herschel} passbands are
still sufficient to undertake this study.
The data fit this simple model of a modified blackbody function well.
A representative SED and its fit are shown in Figure~\ref{fig:sed}.

The \IRAS\ 100~\micron\ brightness was used in fitting SEDs in
previous work \citep{abergelDd2011,Martin2012}, but was not included 
here because of its relatively low angular resolution ($\sim4$\farcm3).
As discussed in Appendix~\ref{appen:iris}, we subsequently
checked that the 100~\micron\ emission is consistent with the SED fit.  
On the other hand, the
70~\micron\ emission is greater than that predicted by the equilibrium
emission SED, confirming the additional contribution from VSGs (see
example in Figure~\ref{fig:sed}).

Errors on the parameters from the SED fit were estimated using Monte
Carlo simulations after assessing the various sources of error in the
intensities of the \her\ images (see Appendix~\ref{sec:errors}).  A
typical SED fit has a $\chi^2$ near two (as expected for four
intensities, two parameters, and thus two degrees of freedom) and the
map of $\chi^2$ is featureless.

\begin{figure*}
\plottwo{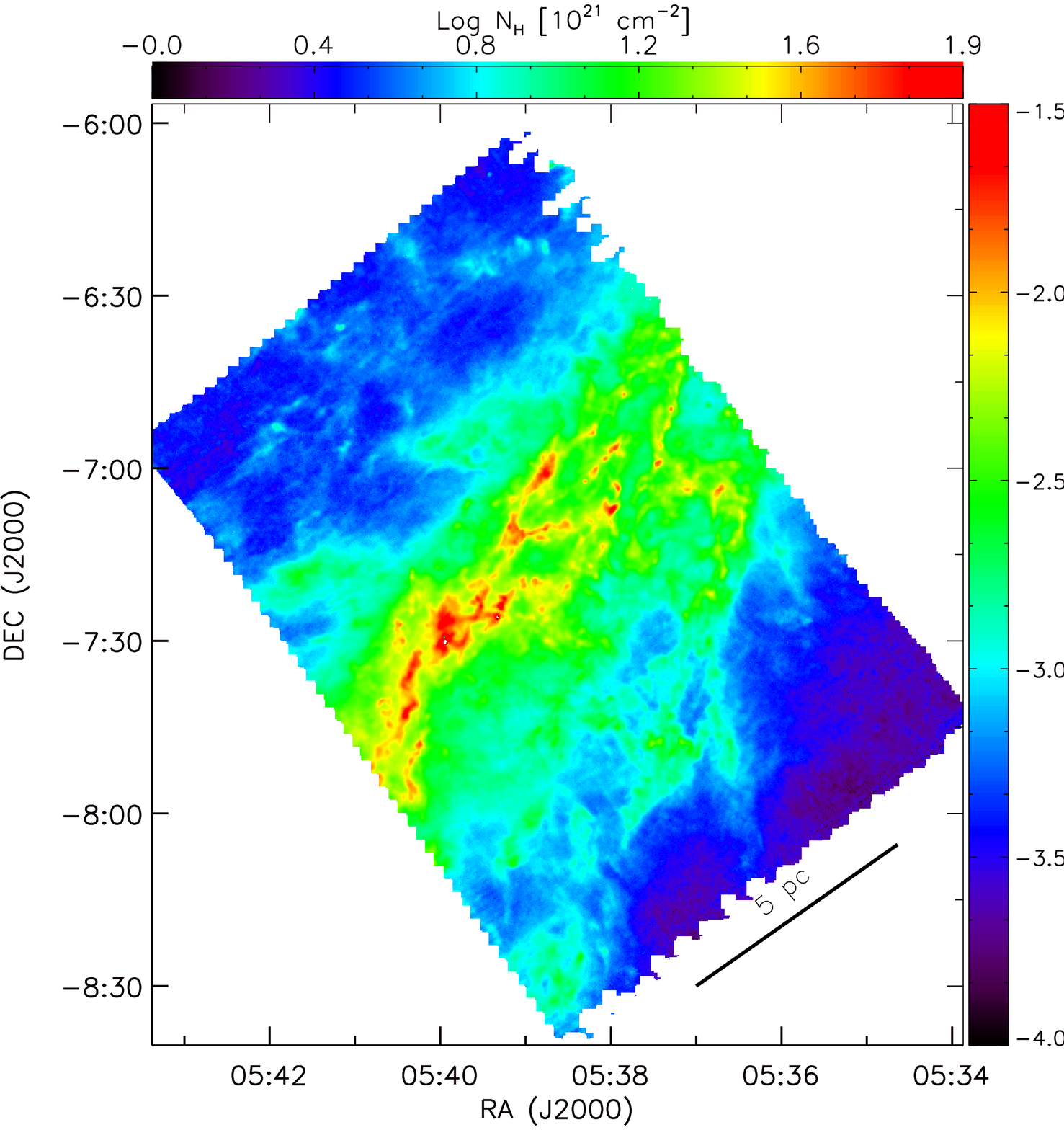}{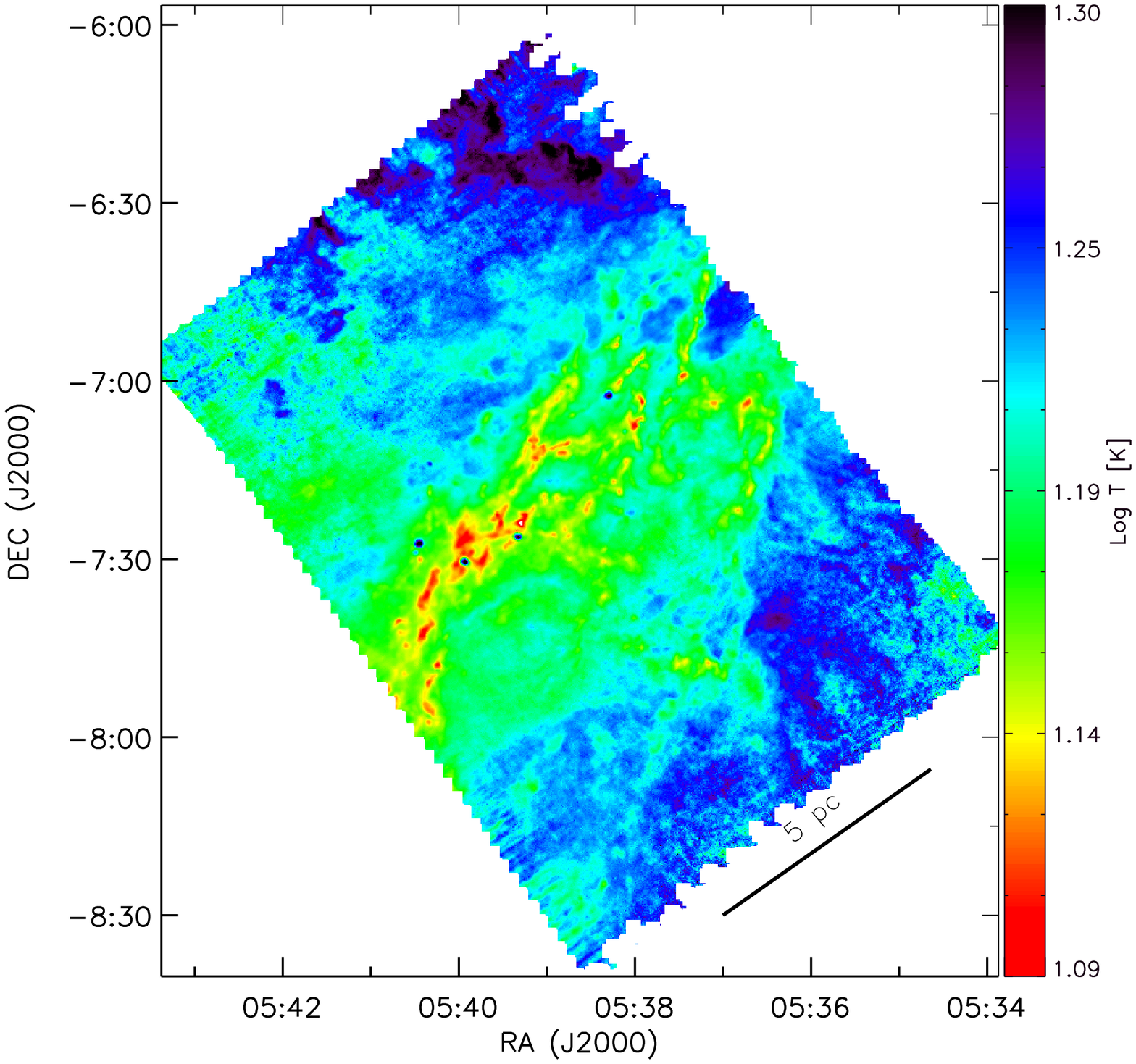}
\caption{\emph{Left}: Optical depth map for Orion~A-C1 as obtained
  from fitting SEDs to the 160-\micron\ PACS and three SPIRE
  intensities, from images brought to a common resolution of
  37\farcs0.  The dust emissivity index $\beta$ was fixed at 1.8.
\emph{Right}: Corresponding temperature map, with a complementary
\emph{inverted} color scale.  The logarithmic range spans 12.3 to 20.0~K.
Note by comparison with the optical depth how the dust in regions of
high column density tends to be cooler ($< 1.14$~dex or 13.8~K) than in
diffuse low column density regions ($> 1.23$~dex or 17~K).
Within the cool filaments are a few compact regions where the dust is
heated by embedded protostars.
The few more diffuse regions with low column density but even higher
dust temperature have an enhanced local radiation field; see also
Figure~\ref{fig:power} in Section~\ref{sec:pspatial} below.}
\label{fig:taut}
\end{figure*}

Figure~\ref{fig:taut} shows results from the SED fits as maps of the
fitting parameters for the Orion~A-C1 field.  The left panel shows a
map of $\tau_{1200}$; the variation of \nh\ is shown by the top color
bar, adopting the mean relation found below (Figure~\ref{fig:nhvsav}).
The corresponding map of $T$ is in the right panel; a complementary
\emph{inverted} logarithmic scale has been used to bring out both the
general anti-correlation (see also Figure~\ref{fig:tvstau} in
Section~\ref{sec:ttau}) and exceptions.  One common feature in these
two maps of the fitting parameters is the filamentary structure, which
is of high column density and generally low temperature, as might be
expected for an IRSF that is attenuated in those regions.  This
behavior is discussed further in Section~\ref{sec:power}.  Filaments
are common in the HGBS
\citep{andre2010,mensch2010,arzou2011,peretto2012}, HOBYS
\citep{motte2010,hill2011,hennemann2012,schneider2012}, and Hi-GAL
\citep{molinari2010} images.


\section{Submillimeter Optical Depth, \nh, and Opacity} \label{sec:results}

\subsection{$\tau$ and \nh} \label{sec:tautnh}

\begin{figure}
\includegraphics[width=\linewidth]{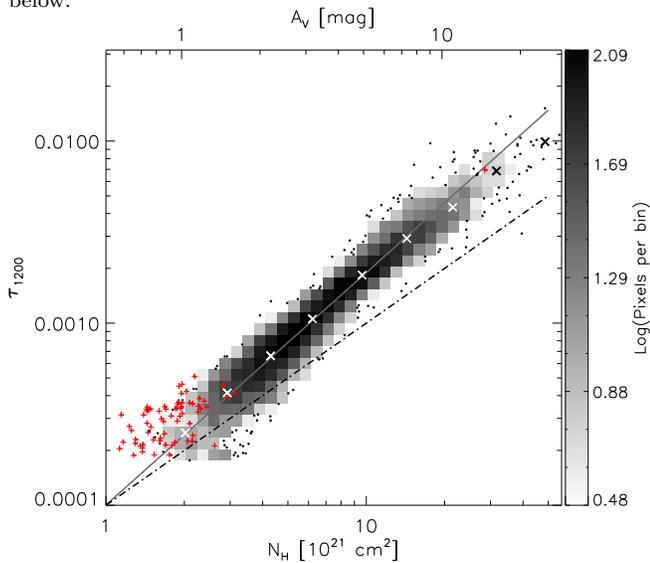}
\caption{$\tau_{1200}$ from SED fits versus column density \nh\ obtained 
from near-infrared color excess \Ejk.
Given the large number of pixels in the maps, we have used a logarithmic 
grey-scale to represent different values in a two-dimensional histogram 
of the number of pixels per bin. 
Individual data points are plotted (dots) where the density of points
is low, less than or equal to three. Data where \nh\ has $S/N <2$ are
plotted as red pluses and are not included in the histogram or fit;
these are also identified in subsequent figures.
The crosses show data binned logarithmically along \nh.
The dotted-dashed line shows a linear relationship between
$\tau_{1200}$ and the underlying \nh\ based on a fixed dust opacity,
here $\sigma_{\rm e}(1200)=1\,\times 10^{-25}$ cm$^2$ H$^{-1}$ typical
of high latitude dust.
Compared to this relation there appears to be a non-linear increase in
$\tau_{1200}$ (and thus \sige(1200)) with \nh.
The solid line is the power-law correlation obtained from fitting
$\tau$ and \nh, excluding the low $S/N$ data. The fit yields a power
law index $1.28 \pm 0.01$ (slope of the line) and amplitude $(1.01 \pm
0.02) \times 10^{-4}$ for the adopted units (intercept at $N_{\rm {H}}
= 10^{21}$~cm$^{-2}$).  }
\label{fig:nhvsav}
\end{figure}

Before comparison with our \Ejk\ proxy for \nh, all parameters
resulting from the SED fits on the \her\ maps were sampled to match
the map of \Ejk\ (Appendix~\ref{samplingtau}).  
Figure~\ref{fig:nhvsav} shows a plot of the sampled $\tau_{1200}$
against \nh\ from the conversion from \Ejk\ in
Equation~(\ref{eq:nhvsejk}).

The two independent measures of column density are well correlated.
The data have some dispersion, with possible reasons including
uncertainty in the measurement of color excess, cosmic scatter about
Equation~(\ref{eq:nhvsejk}) converting \Ejk\ to \nh, and errors in the
determination of $\tau_{1200}$ from the SED.

Observationally, the properties of dust grains have been best
constrained for the high latitude diffuse ISM. For example, the
equilibrium dust temperature is typically near 18~K and \sige(1200) is
$1.0 \times 10^{-25} {\rm cm^{-2}}~ \rm H^{-1} $
\citep{boulanger1996,abergelDd2011}.  The expected linear relation for
this opacity is plotted with the dotted-dashed line in
Figure~\ref{fig:nhvsav}.  This relation is close to that of the
present data at low column density.  However, there appears to be a
significant non-linearity in the data.  The formal fit, excluding \nh\
data with $S/N < 2$, is $\tau_{1200} \propto N_{\rm H}^{1.28 \pm
  0.01}$; the systematic error on the index is 0.03.  Such
non-linearity appears to be evidence for grain evolution.

Recall that $\tau_{1200}$ was derived under the assumption of a
constant $\beta$.  We also derived $\tau_{1200}$ treating $\beta$ as a
(third) free parameter, in case $\beta$ evolves too.  However, we
found that $\beta$ does not change systematically with column density,
and so the same non-linear trend is observed in
Figure~\ref{fig:nhvsav} -- the evidence for grain evolution remains.
There is increased scatter about the mean correlation and the
intercept is 10\% higher because the average $\beta$ is 1.96 (with
higher $\beta$, SED fits tend to have lower $T$ and hence higher
$\tau$).

Although the conversion factor from \Ejk\ to \nh\ might not be
constant for the higher \nh\ in Figure~\ref{fig:nhvsav}, there is
already some divergent non-linear behavior within the range \nh $< 5
\times 10^{21}$~cm$^{-2}$ that is calibrated \citep{Martin2012} .
Nevertheless, we have to be aware that \emph{apparent} non-linearity
might simply be a reflex of an erroneous conversion of \Ejk\ to \nh\
(see Equation~(\ref{eq:nhvsejk})).
To remove the evidence for grain evolution, i.e., to maintain a
constant submillimeter opacity, would require $N_{\rm H} \propto
E(J-K_{\rm s})^{1.28}$ -- which would itself be evidence for grain
evolution, albeit as it affects the near-infrared extinction.  This
dependence, is however, opposite to what has been predicted for the
initial changes in extinction curves resulting from grain evolution by
ice-mantle formation and aggregation \citep{ormel2011}.

Furthermore, empirically, the ratio of \Ejh/\Ehk\ is constant with
column density (Appendix~\ref{sampling}).  \citet{chapman2009} find
the same constancy in the shape of the near-infrared extinction curve
in probes of cores to slightly higher peak optical depth than reached
here, but interestingly they find an increase in the relative amount
of mid-infrared extinction at high optical depth.  A constant shape of
the near-infrared extinction curve, despite changes in the visible
(e.g., as encoded by $R_V$), would be consistent with aggregation of
the smallest interstellar grains with the largest, but not with
aggregation among the largest particles, because in the latter case
the scattering would be greatly enhanced contributing to a steeper
wavelength dependence \citep{kim1994}.  The enhanced mid-infrared
extinction seen at high column densities might be explained by the
addition of ice mantles \citep{chapman2009}.  Possibly the same
affects the submillimeter.
Grain evolution, when and where it occurs, might be complex and
different than this and so it is hard to be definitive.  Until there
is evidence to the contrary, we think it is reasonable to adopt \Ejk\
as a measure of \nh\ over the range of column densities found in this
field, and thus we conclude that it is the submillimeter opacity that
has changed.  But it remains important to understand how the
submillimeter opacity can change without a noticeable change in the
near infrared.

Dense filaments appear to have a characteristic linear width
$\sim0.1$~pc \citep{arzou2011,fischera2012} which is 50\arcsec\ at the
$ \sim400$~pc distance of Orion.  This scale is not much larger than
the 37\arcsec\ resolution of our \her\ maps of $\tau_{1200}$ but is
significantly smaller than the $\sim 3\farcm5$ resolution of the \Ejk\
map (Appendix~\ref{resolution}) required for our opacity study.
Therefore, we do not have enough spatial resolution to investigate in
detail the opacity of dust inside individual filaments (or prestellar
cores) where the column density and volume density would be even
higher.  We are not able to see if the trend indicating grain
evolution continues to even higher densities.
But see the brief discussion ending the next subsection.

The highest column densities ($A_{\rm V} > 8$) are the most likely to
have been influenced by self gravity.  In principle, if extremely
compact structures developed there and had a low covering factor they
might be missed in the `AvMAP' assessment of \Ejk\ because of the
relatively sparse sampling by the 2MASS stars
(Appendix~\ref{sampling}), and yet they would still contribute to the
\her\ assessment of $\tau_{1200}$ because their emission would be
included in the finite beam.  If that were the case \emph{and} if such
structures had a significant column density, then the \nh\ derived
from \Ejk\ would be underestimated relative to the sampled
$\tau_{1200}$, and so the observed trend in Figure~\ref{fig:nhvsav}
would turn up above this high column density.  If anything, the data
there fall slightly below the non-linear dependence shown.  This
mapped region of the molecular cloud does have some compact
(self-gravitating and probably quasi-equilibrium) high column density
structures, filaments and even some `cores,' with enough mass to be
detectable in the submillimeter, but their physical size is large
enough that at the distance to the Orion A cloud these structures are
measured to be resolved, and so are not as in our hypothetical
scenario physically vanishingly small and not sampled by the 2MASS
stars.
Another constraint on any hypothesis aiming to account for, or explain
away, this non-linearity is that the trend in Figure~\ref{fig:nhvsav}
begins at quite low column densities.  Furthermore, there is ample
evidence for a higher opacity associated with high column density
molecular regions (see, e.g., the summary in Appendix A in
\citealp{Martin2012}, and the following subsection).


\subsection{Opacity  $\sigma_{\rm e}(1200)$ and \nh}\label{sec:sigma}

\begin{figure}
\includegraphics[width=\linewidth]{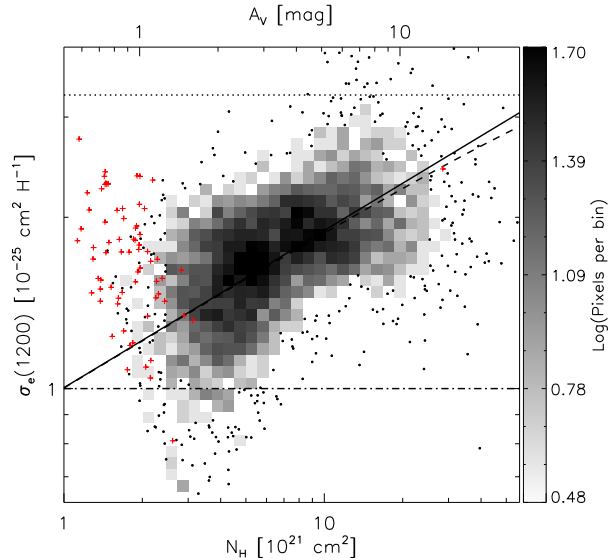}
\caption{Opacity \sige(1200) versus \nh. The solid line shows a
power-law variation \sige(1200) $\propto$ \nh$^{0.28}$
taken from the solid line in Figure~\ref{fig:nhvsav}. The dashed line
depicts the effect of color correction due to the finite bandwidth of
the 2MASS filters on the measurement of \Ejk\ (column density).
   The horizontal dotted-dashed line shows the average \sige(1200)
   for the diffuse ISM. The dotted line is for
   $\sigma_{\rm e}(1200)=3.3\, \times 10^{-25}$ cm$^2$ H$^{-1}$, the
   standard adopted by the HGBS and HOBYS. Pixels representing 
   two-dimensional histogram is same as described in Figure \ref{fig:nhvsav}.
  Individual data points are plotted where the density of points is 
   less than or equal to three.  }
\label{fig:sig}
\end{figure}
 
The nature of the systematic change in \sige(1200) is revealed more
explicitly in Figure~\ref{fig:sig}, which is just an alternative
representation of Figure~\ref{fig:nhvsav}, from the sampled
$\tau_{1200}$ map simply divided by the \nh\ map, plotted on an
expanded vertical scale.  The solid and dotted-dashed lines are the
same as defined for Figure \ref{fig:nhvsav}.  Despite the dispersion
about the solid line, the overall trend clearly indicates an increase
of $\sigma_{\rm e}(1200)$ with column density ($\sigma_{\rm e}(1200)
\propto N_{\rm H}^{0.28 \pm 0.01 \pm 0.03}$), by at least a factor of two over
this \nh\ range.

A particular additional perspective provided by the present study of
this region in Orion~A is that there ought to be a rough
correspondence between volume density and column density. This link
could enable a more direct connection of grain evolution to
environment.
Compared to the density of the diffuse ISM ($n_{\rm H} \sim 1$ --
10~cm$^{-3}$), the density typical of a star formation region where
clumps are beginning to show signs of gravitational contraction is
several orders of magnitude higher (10$^3$ -- 10$^5$ cm$^{-3}$).
In the present analysis, we are probing environments, especially in
filamentary structures, with $n_{\rm H}$ up to $\sim10^4$~cm$^{-3}$.

The measurement of \sige(1200) is sensitive to the estimate of
temperature through the Planck function; see
Equation~(\ref{emissmass}) where we have assumed a constant $T$ along
the line of sight.  Obviously, either for passively-heated molecular
clouds with significant optical attenuation or for those with an
internal energy source, there is a gradient in temperature which we
have ignored.  Particularly relevant here is the former possibility
which has been studied extensively using radiative transfer
simulations. For example, \citet{Ysard2012} found that the temperature
obtained from fitting a single-temperature SED is overestimated
compared to the true dust temperature for central lines of sight
through filaments or spherical clouds.  The difference increases
non-linearly with central column density and is noticeable by $N_{\rm
  {H}} \sim 14 \times 10^{21}$~cm$^{-2}$, depending on the grain
model.  In the context of our analysis, an `overestimate' of the
temperature would have suppressed the optical depth derived here;
hence, this effect would enhance rather than erase the trend of
increasing \sige(1200) with \nh.

The dashed line in Figure~\ref{fig:sig} shows the correction of the
correlation line when the effect of the finite filter bandwidths on
the color excess is taken into account (see
Appendix~\ref{sec:band-width}).  Underestimation of \Ejk\ makes the
rate of increase in \sige(1200) only marginally smaller.

For the pixels with low column density, the derived values of
\sige(1200) are close to the value $1 \times 10^{-25}$ cm$^2$ H$^{-1}$
for high latitude diffuse interstellar dust \citep{boulanger1996} and
typically within the range of $(0.6 - 1.6) \times 10^{-25}$ cm$^2$
H$^{-1}$ found by \citet{abergelDd2011}.
We find that \sige(1200) increases systematically with \nh\
by at least a factor two.
Such high opacities are not unprecedented, one example
being $3.8 \times 10^{-25}$ cm$^2$ H$^{-1}$ for a dense molecular
region in the Vela molecular ridge where the column density ranged
between 10 to $40 \times10^{21}$ cm$^{-2}$ \citep{Martin2012}.
For a molecular region in Taurus of intermediate column density
($\sim10^{22}$ cm$^{-2}$), \citet{abergelTd2011} obtained an opacity of
$(2.3 \pm 0.3) \times 10^{-25}$~cm$^2$~H$^{-1}$.  By contrast, for the
neighboring atomic phase with $N_{\rm {H}} < 3 \times
10^{21}$~cm$^{-2}$, they obtained $(1.14 \pm 0.2) \times
10^{-25}$~cm$^2$~H$^{-1}$.
Based on a complementary new method, physical modeling of the
brightness profiles of filaments, \citet{fischeraestimates2012} find a
range $(1.3 - 2.8) \times 10^{-25}$~cm$^2$~H$^{-1}$ (the range
depending on the adopted distance).
These are consistent with the trend seen in Figure~\ref{fig:sig}.

The standard opacity adopted by the HGBS and HOBYS (e.g.,
\citealp{andre2010,motte2010}) is $\sigma_{\rm e}(1200)=3.3\, \times
10^{-25}$ cm$^2$ H$^{-1}$ (see Figure~\ref{fig:sig}).  This is taken
from theoretical calculations by \citet{preibisch1993} for evolved
dust in protostellar cores (see also \citealp{ossenkopf1994}).
Although we do not probe to such high densities, this value seems a
reasonable extrapolation of the trend in Figures~\ref{fig:nhvsav} and
\ref{fig:sig}; a corollary is that a single value of the opacity
cannot be used for all environments across a region.

As emphasized by \citet{shirley2011}, the largest uncertainty in
determining the mass of a core is in the adopted value of the opacity.
Their modeling of the extended submillimeter emission in the cold
envelope of B335, combined with deep imaging at $H$ and $K$ ($E(H-K)$
up to 3, about three times larger than in our map), allowed them to
constrain the ratio of the submillimeter opacity, at 850~\micron, to
the opacity (scattering plus absorption) at 2.2~\micron.  They found a
range $(3.2 - 4.8) \times 10^{-4}$.  Extrapolating this to the ratio
involving the opacity at 250~\micron\ using their estimate of $\beta$
(2.18 to 2.58) gives a range $(4.6 - 11.3) \times 10^{-3}$.  This can
be compared to our estimate of the same ratio, $(2.3 - 4.9) \times
10^{-3}$ for $\sigma_{\rm e}(1200)= (1 - 2) \times 10^{-25}$ cm$^2$
H$^{-1}$ (note that this ratio is independent of the calibration of
either opacity to \nh).  Thus regions characterized by higher density,
or higher column density here, both appear to have an enhanced
submillimeter opacity relative to that in the $K$ band (and one cannot
rule out that both might have increased).  Of course, as we have noted
in the Introduction, what is really needed to find the mass is the
product $r \kappa_\nu$, and so without some direct calibration as
attempted here, the Achilles heel is to first derive the submillimeter
dust opacity itself from the above opacity ratio and then adopt an
appropriate dust to gas ratio $r$.

%


\section{Power} \label{sec:power}

Big dust grains bathed in the ISRF absorb energy at higher frequencies,
and in thermal equilibrium they re-emit the same amount of energy
at much lower frequencies.  Thus, the observed $P$ is what the grains were
able to absorb in their environment, per H.  For the dust emissivity
relation ($\beta$) adopted, $P$ can be expressed as
\begin{equation}
P/P_0 = (\sigma_{\rm e}(1200)/\sigma_0)\, (T/T_0)^{5.8},
\label{eq:psigt}
\end{equation}
normalized in terms of the average parameters in the diffuse atomic
ISM at high latitude
\citep{abergelDd2011}: $T_0 = 17.9$~K, $\sigma_0 = 1.0 \times
10^{-25}$~cm$^2$~H$^{-1}$ and $P_0 = 3.8 \times 10^{-31}$~W~H$^{-1}$.


\subsection{$T$ and $\tau$}\label{sec:ttau}

\begin{figure}
\includegraphics[width=\linewidth]{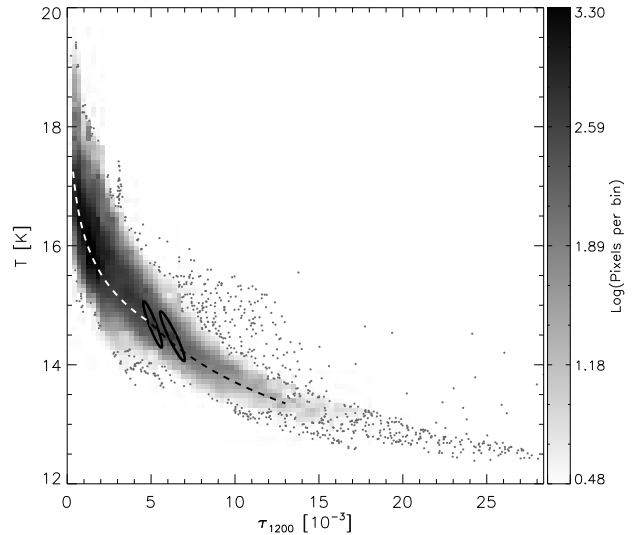}
\caption{$T$ versus $\tau_{1200}$ from the SED fits, with the
logarithmic grey-scale representing the values in a two-dimensional
histogram.
Representative error ellipses for a single pixel are from Monte Carlo simulations of the
SED fitting.
The dashed curves are power laws schematically following the
anti-correlation. Individual data points are plotted where the density
of points is less than or equal to three.}
\label{fig:tvstau}
\end{figure}

The value of the equilibrium grain temperature is not a primary dust
parameter.
Rather, as is clear from Equation~(\ref{eq:psigt}), $T$ is a parameter
that \emph{responds} to environmental changes in the ISRF affecting
the absorbed $P$ and to changes in the grain properties like
\sige(1200) and possibly the absorption which also affects the
absorbed $P$.

The anti-correlation of $T$ and $\tau_{1200}$ seen in the maps of
Figure~\ref{fig:taut} and similarly for the Orion~A-S1 field is
highlighted in Figure~\ref{fig:tvstau}.
Because we are primarily interested in regions where there is no
internal source of energy, we have not plotted data for the pixels
corresponding to a few identified embedded protostars (high $T$ and
intermediate $\tau_{1200}$).
The dashed curves are not a fit but rather trace the general trend
schematically.  We have transferred these loci to
Figures~\ref{fig:powervsav} - \ref{fig:pvssig} below to gain insight
into relationships between other variables.
For simplicity, we have used two power laws, $T/T_{\rm b} =
(\tau(1200)/\tau_{\rm b})^{-\alpha}$, where a break occurs at $T_{\rm
  b} = 14.7$~K and $\tau_{\rm b} = 4.9 \times 10^{-3}$, and where
$\alpha = 0.06$ and 0.1 below and above the break, respectively.


\subsection{Spatial Dependence of $P$}\label{sec:pspatial}

\begin{figure}
\includegraphics[width=\linewidth]{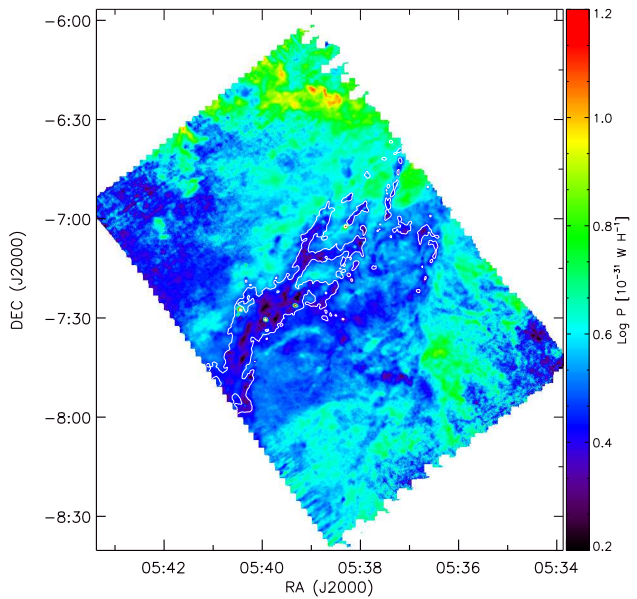}
\caption{Map of the specific power $P$ in the Orion~A-C1 field.  The
overplotted white contour, corresponding to $N_{\rm H} = 20
\times10^{21}$~cm$^{-2}$ (\av\ $\sim10.6$ mag), reveals how $P$
decreases significantly in high column density regions, especially
here along filaments.}
\label{fig:power}
\end{figure}

We produced a map of $P$ shown in Figure~\ref{fig:power} making use of
the $T$ and $\tau$ maps (as in Figure~\ref{fig:taut}) and the average
dependence of \sige(1200) on $\tau_{1200}$ from
Figures~\ref{fig:nhvsav} and \ref{fig:sig}.
More directly from the data, a map of the radiated energy is obtained
from the integration of the respective SEDs over all frequencies.  As
in Section~\ref{sec:sigma}, we divide the resulting sampled map by the
\nh\ map to obtain a map of the specific power $P$
(Equation~(\ref{power})).  This is very similar to
Figure~\ref{fig:power}.\footnote{The displayed map of $P$ is shown at
  higher spatial resolution than a map of $\sigma_{\rm e}(1200)$ would
  have to highlight the effects of attenuation of high frequency
  radiation in the filamentary structures.}

For the pixels corresponding to low \nh, $P$ is on average about
3.7$\times 10^{-31}$ W~H$^{-1}$, close to the value $3.8 \times
10^{-31}$ W~H$^{-1}$ for the typical specific power in the diffuse ISM
at high Galactic latitude \citep{abergelDd2011}.
In a few diffuse regions in our maps, for example the northern part of
Orion~A-C1, $P$ gets as high as $10 \times 10^{-31}$ W~H$^{-1}$,
clearly showing an enhancement of the local radiation field by a factor
of $\sim2.5$ times relative to the average ISRF.
Others examples are the B77 Bright Nebula and a halo around V1792 Ori,
both in the Orion~A-S1 field and so not shown here.
These regions also have strong emission at 70~\micron\ and 
160~\micron\ in the respective \her\ images.

As illustrated in Figure~\ref{fig:power} by the overlay of the contour
$N_{\rm H} = 20 \times 10^{21}$~cm$^{-2}$, regions with the highest
\nh\ tend to have lower $P$.  The most straightforward qualitative
interpretation is in terms of attenuation of the ISRF.  The high
frequency components of the ISRF that are primarily responsible for
the heating of the dust grains cannot penetrate easily into the
interior of the denser molecular structures.  Most of the power is
absorbed within the outer layers of the cloud and so the dust in the
core generally has a lower equilibrium temperature.

It should be noted that only a fraction of the total \nh\ through the
center of the filaments is associated directly with the structure, the
rest arising from the embedding and foreground and background
material.  Therefore, the detailed relationship between specific power
and column density requires knowledge of density profiles and cloud
geometry, the strength of the ambient ISRF, and detailed radiative
transfer modeling taking into account appropriate dust populations and
absorption and scattering properties \citep{Fischera2011,Ysard2012}.

In contrast, and not surprisingly, close inspection of
Figure~\ref{fig:power} reveals a few lines of sight where there is an
internal source of energy (protostar(s)) within a dense compact
structure. With this extra energy input, the resulting $P$ is higher
locally.  These are readily seen in Figure~\ref{fig:taut}~(right) as
warmer than the immediate surroundings.  An example is at $RA, Dec$
05:39:56, $-$7:30:27, in a complex environment with the brightest hot
source named LDN 1641 S3 IRS/ FIRSSE 101.


\subsection{$P$ and \nh}\label{sec:powernh}

\begin{figure}
\includegraphics[width=\linewidth]{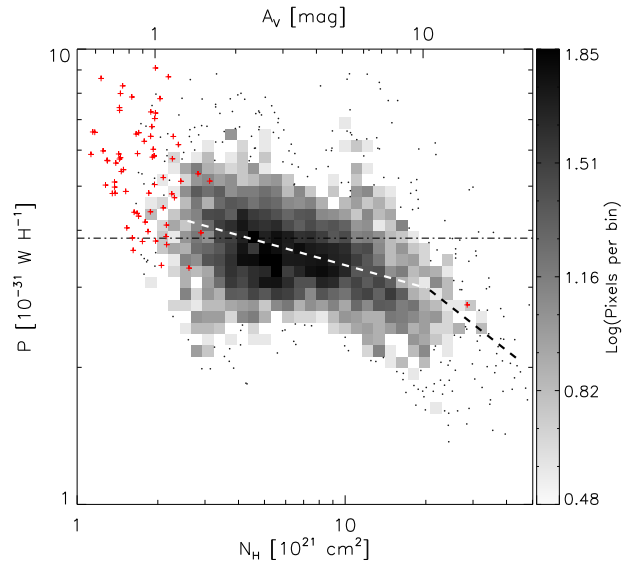}
\caption{Specific power $P$ versus column density \nh.  $P$ fluctuates
  about the typical high latitude diffuse ISM value of 3.8 $\times 10^{-31}$ W H$^{-1}$
  (dotted-dashed line).  For
  high column density, $P$ decreases because of attenuation of the
  ISRF.
Dashed lines show the loci transferred combining trends from
Figures~\ref{fig:nhvsav} - \ref{fig:tvstau}. 
Pixels representing two-dimensional histogram is same as described in Figure 
\ref{fig:nhvsav}.
Individual data points are plotted where the density of points is less than or equal to
three.
}
\label{fig:powervsav}
\end{figure}

Figure~\ref{fig:powervsav} shows the relationship of sampled $P$ to \nh.
There is a trend of decreasing $P$ at higher column density that sets
in well before the value $20 \times 10^{21}$~cm$^{-2}$ emphasized
in Figure~\ref{fig:power}.
Above we quantified trends relating \sige(1200) and $\tau_{1200}$ to
\nh, and $T$ to $\tau_{1200}$.  If we substitute these trends in
Equation~(\ref{eq:psigt}), we find $P/P_{\rm b} = (N_{\rm H}/N_{{\rm
    H}_b})^{-\delta}$, where $P_{\rm b} = 2.98 \times 10^{-31}$
W~H$^{-1}$ and $N_{{\rm H}_{\rm b}} = 20.3 \times 10^{21}$~cm$^{-2}$,
and where $\delta = 0.165$ and 0.46 below and above the break,
respectively.  These loci are plotted in Figure \ref{fig:powervsav}.


\section{Interrelationships among $T$, $P$, and \sige(1200)}\label{sec:inter}

Among $T$, $P$, and \sige(1200), a scatter plot of any two provides 
insight into their interrelationship.  On the resulting  figures,
 we also plot loci defined by fixing the third quantity.


\subsection{$T$ -- \sige(1200) Relation}\label{sstsig}

\begin{figure}
\includegraphics[width=\linewidth]{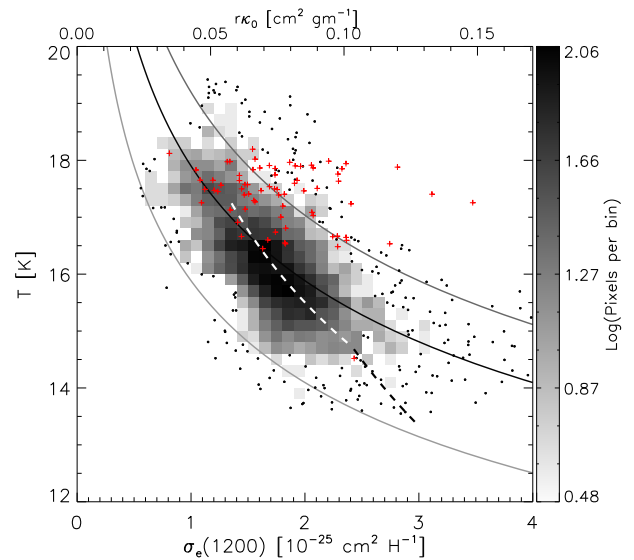}
\caption{Dust temperature $T$ versus dust opacity
\sige(1200).  Solid lines are the loci of constant specific power,
$P$. The middle line is for 3.8 $\times 10^{-31}$ W~H$^{-1}$, the
typical $P$ in the high latitude mean ISRF (see Section~\ref{sec:power}).
The lower and upper  lines are for an ISRF different by  factors of
0.5 and 1.5, respectively.
Dashed lines show the loci transferred combining trends from
Figures~\ref{fig:nhvsav} - \ref{fig:tvstau}.  Pixels representing 
two-dimensional histogram is same as described in Figure \ref{fig:nhvsav}. 
Individual data points 
are plotted where the density of points is less than or equal to three.
}
\label{fig:tvssig}
\end{figure}

As discussed above, the dust temperature is determined by the energy
balance between the power absorbed from the ISRF and emission.  For a
given intensity of the ISRF and absorption coefficient, and thus
absorbed power, grains with a higher \sige(1200) radiate more
efficiently and so attain a lower equilibrium temperature
(Equation~\ref{eq:psigt}).  Therefore, the observed equilibrium dust
temperature and the intrinsic dust property
\sige(1200)\footnote{\sige(1200) also involves the dust to gas ratio.}
are expected to be inversely related.  In Figure~\ref{fig:tvssig}, we
show the anti-correlation between sampled $T$ and \sige(1200) for the
Orion~A region.  The solid line shows a locus of constant specific
power $P=3.8\times 10^{-31}$ W~H$^{-1}$ whereas the loci below and
above are 0.5 and 1.5 times this value, respectively.  There is a
range of $P$ and indeed, unlike in the high latitude diffuse ISM, we
do not expect to have constant $P$ in the regions of high column and
volume density (see Figure~\ref{fig:powervsav}), due to attenuation by
dust itself.  To capture this effect, we have transferred the loci
from the above figures.
We find $T/T_{\rm b} = (\sigma_e(1200)/\sigma_b)^{-\epsilon}$, where
$\sigma_{\rm b} = 2.41 \times10^{-25}$~cm$^2$~H$^{-1}$ and where
$\epsilon = 0.27$ and 0.46 below and above the break, respectively.
These loci are plotted in Figure~\ref{fig:tvssig}.


\subsection{$P$ -- \sige(1200) Relation}\label{sspsig}

\begin{figure}
\includegraphics[width=\linewidth]{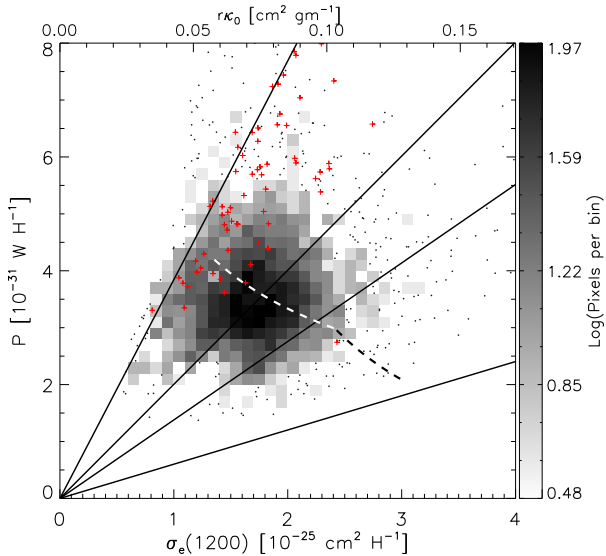}
\caption{Specific power $P$ versus dust opacity
  \sige(1200).  The solid lines (bottom to top) are the loci of
  constant temperatures equal to 13, 14, 15, and 17.9~K.
Dashed lines show the loci transferred combining trends from
Figures~\ref{fig:nhvsav} - \ref{fig:tvstau}.  
Pixels representing  two-dimensional histogram is same as described 
in Figure \ref{fig:nhvsav}.
Individual data points
are plotted where the density of points is less than or equal to three.}
\label{fig:pvssig}
\end{figure}

Figure \ref{fig:pvssig} shows sampled $P$ versus \sige(1200), and is
sufficient to complete our examination of the three-parameter
relations of Equation~(\ref{eq:psigt}) ($P$ versus $T$ being
redundant).
For a given dust equilibrium temperature, the power radiated in the
submillimeter is proportional to the emission cross-section.  The
solid lines are the loci of constant temperature with increasing slope
corresponding to 13, 14, 15, and 17.9~K, respectively.
For dust grains exposed to same ISRF (constant $P$), the less
efficient emitters (lower \sige(1200)) will have higher temperature.

To capture the effect of attenuation, we have transferred the loci from
the previous figures.
We find $P/P_{\rm b} = (\sigma_{\rm e}(1200)/\sigma_{\rm b})^{-\iota}$, where
$\iota = 0.59$ and 1.65 below and above the break, respectively.
These loci are plotted in Figure~\ref{fig:pvssig}.

 
\section{Conclusion}\label{sec:conclusion}

We have studied the properties of dust grains using multi-wavelength
images of thermal dust emission in the Orion~A molecular cloud at 160,
250, 350, and 500~\micron, acquired by the PACS and SPIRE cameras on
\her\ as part of the HGBS.  We fit a modified blackbody model to the
spectral dependence of the surface brightness of each pixel.  Although
dust along the line of sight is probably not all at the same
temperature, assuming a single $T$ is a reasonable model providing
good quality SED fits to the \her\ data.  Thus, we obtained the
spatial distribution of $T$ and $\tau_{1200}$ across the map.

The derived optical depth map is well correlated with the \nh\ map
derived from \Ejk\ and we combined them to explore how the
dust opacity \sige(1200) varies with column density.
A submillimeter opacity \sige(1200) near $1 \times
10^{-25}$~cm$^{2}$~H$^{-1}$ is typical for diffuse dust along high
latitude lines of sight of low column density (\av $\sim$ 1 mag) and
close to that found at low column density in these Orion~A fields.  On
lines of sight intercepting high column density filamentary
structures, however, we found evidence for an increase of \sige(1200)
by a factor two or more, approaching that adopted by the HGBS and
HOBYS.  Overall, there is a systematic trend $\sigma_{\rm e}(1200)
\propto N_{\rm H}^{0.28}$.  This dependence is strong evidence for
grain evolution with environment.
There is not a single opacity that can be applied to a whole mapped
region with environments ranging from diffuse to higher column density
structures.  This has quantitative implications for the interpretation
of the probability distribution function of the column density and
possibly for the shape of the core mass function.

For the low column density lines of sight, the average value for the
specific power $P$ is $3.7 \times10^{-31}$~W~H$^{-1}$, equivalent to
1.2~\lsol/\msol.
The power emitted is equal to the power absorbed and so depends on the
local ambient ISRF.
Thus, the decrease of $P$ that we see in the dense filamentary
structures ($P \sim 1.5\times10^{-31}$~W~H$^{-1}$) can be attributed
to attenuation of the ISRF.  There are also local enhancements of the
ISRF.
Overall, $P$ ranges over a factor of 10, roughly centered on the diffuse
ISM value.

The emission opacity \sige(1200) of big grains together with the
relative strength of the ISRF determines the equilibrium temperature.
In the diffuse ISM where dust grains are exposed fully to the ISRF, the
temperature is typically 18~K.  Inside dense filamentary structures in
the observed region of the Orion~A molecular cloud, the opacity is
larger and the ISRF is attenuated, both leading to dust temperatures
lower than 14~K.

In this high latitude region in the Orion~A molecular cloud, high
column density arguably corresponds to high volume density.  The
observed change in the optical properties of dust grains in dense cold
regions can be attributed to grain growth due to aggregation and/or
accretion of ice mantles.
This process might reasonably be expected to correlate with a decrease
in the relative abundance of VSGs. It is difficult, however, to probe
the presence of the latter because the radiation that would normally
excite their non-equilibrium emission at shorter wavelengths ($<$
100~\micron) is sharply attenuated in these very regions.


\acknowledgements 

We thank the referee for helpful comments re clarification of the
results in Section~\ref{sec:results}.  This research was supported in
part by the Canadian Space Agency (CSA) and the Natural Sciences and
Engineering Research Council of Canada.
SPIRE has been developed by a consortium of institutes led by Cardiff
Univ. (UK) and including: Univ. Lethbridge (Canada); NAOC (China);
CEA, LAM (France); IFSI, Univ. Padua (Italy); IAC (Spain); Stockholm
Observatory (Sweden); Imperial College London, RAL, UCL-MSSL, UKATC,
Univ. Sussex (UK); and Caltech, JPL, NHSC, Univ. Colorado (USA). This
development has been supported by national funding agencies: CSA
(Canada); NAOC (China); CEA, CNES, CNRS (France); ASI (Italy); MCINN
(Spain); SNSB (Sweden); STFC, UKSA (UK); and NASA (USA).  
PACS has been developed by a consortium of institutes led by MPE
(Germany) and including UVIE (Austria); KU Leuven, CSL, IMEC
(Belgium); CEA, LAM (France); MPIA (Germany); INAF-IFSI/OAA/OAP/OAT,
LENS, SISSA (Italy); IAC (Spain). This development has been supported
by the funding agencies BMVIT (Austria), ESA-PRODEX (Belgium),
CEA/CNES (France), DLR (Germany), ASI/INAF (Italy), and CICYT/MCYT
(Spain).


\appendix

\section{2MASS Color Excess and Column Density} \label{sec:excess}

Column density can be estimated by measuring extinction or color
excess.  Very productive use has been made of the near-infrared 2MASS
data in the J, H, and ${\rm K_s}$ passbands, post-processed by a
variety of techniques (e.g.,
\citealp{rowles2009,dobashi2011,lombardi2011}).
To stay close to the data and avoid unnecessary additional assumptions
(see also Appendix~\ref{calibration}), here we quantify the extinction
in terms of the near-infrared color excess \Ejk, rather than the total
visual extinction $A_{\rm V}$.  For comparison to values cited in
other literature, the adopted conversion has been $A_{\rm V} = 5.89\,
E(J-K_{\rm s})$.

\subsection{Sampling the Color Excess}\label{sampling}

The extinction map was derived using the `AvMAP' procedure
\citep{Schneider2011} as a weighted average of measures of \Ejh\ and
\Ehk, and subsequently expressed here as \Ejk.  The map is created
from a weighted mean of the color excesses of individual stars.  The
weighting function is a Gaussian of $FWHM = 4\arcmin$ centered on each
$2\arcmin$ pixel and stars are considered out to a radius of $3\,
\sigma$ ($1.3\, FWHM$).  To monitor the quality, maps of the weighted
number of stars used and the weighted standard deviation of the mean
(i.e., the uncertainty of \Ejk) are also computed.

Given the high column densities reached in this field, the issue
arises whether the color excess maps might somehow be `saturated,'
i.e., missing the highest values.  This issue does not seem to be
addressed in the literature.  We have made scatter plots of the maps
of \citet{rowles2009} and \citet{dobashi2011} compared to our \Ejk\
map.  These plots show good correlations with the expected slope up to
$A_{\rm V} \sim 8$, after which these other products cease to
continue to rise as our \Ejk\ grows further.

Maps of \Ejh\ and \Ehk\ can be derived separately.  These
correlate very well with a slope  of $1.73 \pm 0.01$, close to the
`normal' value 1.7 (see discussion in \citealp{Martin2012}).
Like \citet{lombardi2011} have also shown for the Orion~A cloud, this
normal color-color relation extends beyond $E(J - H)= 1.7$~mag and
$E(H - K_{\rm s})= 1$~mag, though the correlation becomes noisier.
Thus, there seems to be no evidence for either the effects of grain
evolution or of saturation of one color with respect to the other up
to $E(J - K_{\rm s})= 2.7$~mag which corresponds to $A_{\rm V} \sim
16$~mag.  Nevertheless, we need to acknowledge the issue and ensure
that the column density from \her\ $\tau$ is sampled on a closely
comparable basis.  This is accomplished as described in
Appendix~\ref{samplingtau}.

\subsection{Gauging the Resolution}\label{resolution}

In our \her\ maps of $\tau_{1200}$ at 37\arcsec\ resolution there is
structure on smaller scales than the $4\arcmin$ weighting function of
`AvMAP.'  Therefore, it is important that the map of $\tau_{1200}$ be
brought to the lower resolution of the \Ejk\ map for an unbiased
comparison (Section~\ref{sec:tautnh}).
Establishing the precise resolution of color excess maps derived from
2MASS data has not been a high priority and so we have done our own
explorations.  A fruitful approach was to use the power spectrum of
the map.  A traditional map (image) produced by an instrument with a
`beam' (or point spread function) is typically oversampled relative to
the beam size and so the information in neighboring pixels is
correlated.  Therefore, there is a predictable roll-over of the power
spectrum at high spatial frequencies corresponding to the size of the
beam.  Although the described Gaussian-weighted and truncated sampling
used in `AvMAP' is not exactly convolution, nonetheless we found an
empirical roll-over of the power spectrum that can be described
quantitatively by a beam of $FWHM$ 3\farcm5, which seems reasonable.

\subsection{Sampling the \her\  $\tau_{1200}$ map}\label{samplingtau}

Ideally, the two independent measures of column density, \Ejk\ and
$\tau_{1200}$, would sample the sky in the same way.  Rather than
trying to bring the higher resolution $\tau_{1200}$ map to the lower
effective resolution of the \Ejk\ map by smoothing,\footnote{In
  preliminary work, the \her\ maps of $\tau_{1200}$ were smoothed and
  then regridded.  For smoothing, we tried both a simple boxcar
  average of 9 by 9 pixels ($11\farcs5$ pixels) and convolution with a
  2\farcm5 Gaussian, finding very similar results.  Compared to these
  degraded $\tau_{1200}$ maps, our `AvMAP'-produced maps of \Ejk\ (see
  also Figure~1 of \citealp{lombardi2011}) are not obviously more
  blurry, i.e., missing all the small-scale high column density
  features, but this assessment is qualitative.}
a more direct approach is to sample the $\tau_{1200}$ map at the
positions of 2MASS stars, forming the weighted average in the same way
as in `AvMAP.'  We found that simply extracting cataloged 2MASS stars
according to the criterion of photometric $S/N > 7$ produced a
weighted number of stars close to `AvMAP.'  While this is not
precisely an identical sampling, nor is the original $\tau_{1200}$
known to pencil-beam precision.  Thus we obtain a map of $\tau_{1200}$
on the same pixels as for \Ejk\ and with comparable sampling of the
column density and resolution pixel by pixel.
Quantitatively, there is a roll-over in the power spectrum of the
sampled $\tau_{1200}$ map for each of the two \her\ fields that
confirms a typical resolution 3\farcm5.

\subsection{Calibration of the Column Density}\label{calibration}

\citet{Martin2012} made a direct calibration of \nh\ in terms of \Ejk:
\begin{equation}
N_{\rm {H}} = (11.5 \pm 0.5) \times 10^{21}\, E(J-K_{\rm s})\, {\rm cm^{-2}}.
\label{eq:nhvsejk}
\end{equation}
We adopted this empirical relation even though it is calibrated up to
only $N_{\rm {H}} \sim 5 \times 10^{21}$~cm$^{-2}$.  The column
density in this field, gauged in the near-infrared, ranges to almost
an order of magnitude larger, but such column densities are not
accessible in the calibration because that depends on measurements of
H and ${\rm H_2}$ column densities made in the ultraviolet.

Instead of \nh, column density from these 2MASS-derived maps is often
expressed in terms of visual extinction, $A_{\rm V}$.  We prefer not
to do so, because we suspect that $A_{\rm V}/E(J-K_{\rm s})$ will vary
in the most dense molecular regions, more so than $N_{\rm
  {H}}/E(J-K_{\rm s})$.  Nevertheless, since this is such standard
practice, for comparison with the literature we provide $A_{\rm V}$ as
well, using the standard diffuse ISM conversion factor; this $A_{\rm
  V}$ is just not to be taken literally.

\subsection{Effect of Finite Bandpasses on the Color Excess}\label{sec:band-width}

Here we evaluate the effect of the finite width of 2MASS photometric
filter response (the $RSRF$; \citealp{cohen2003}) on the observed
extinction.
The attenuated signal for a background star measured through a broad
photometric filter depends upon various parameters such as the
intrinsic SED of the source, the width of the $RSRF$, the shape of the
extinction curve, and the column of material along the line of sight.
Mathematically, the signal can be expressed as
\begin{equation}
F_{\rm filter}(\tau_0)= F_{\nu_0} \int_{\nu_L}^{\nu_U} \left(\frac{\nu}{\nu_0}\right)^{\gamma} e^{-\tau_0 (\nu/\nu_0)^{\alpha}}RSRF(\nu) d \nu,
\end{equation}
where the optical depth is $\tau_0$ at the central frequency $\nu_0$
of the bandpass, $\alpha$ is the exponent of the power-law
approximation to the infrared extinction, $\gamma$ is the spectral
index of the background star across the bandpass, and the limits of
the integration span the bandpass.  The apparent extinction is
therefore $A_{\rm filter} = -2.5 \log (F_{\rm filter}(\tau_0) /F_{\rm
  filter}(0))$.  This value can be compared to the `monochromatic'
extinction $1.086\, \tau_0$.  The J band is most affected because of
its higher extinction.

\begin{figure}
\includegraphics[width=\linewidth]{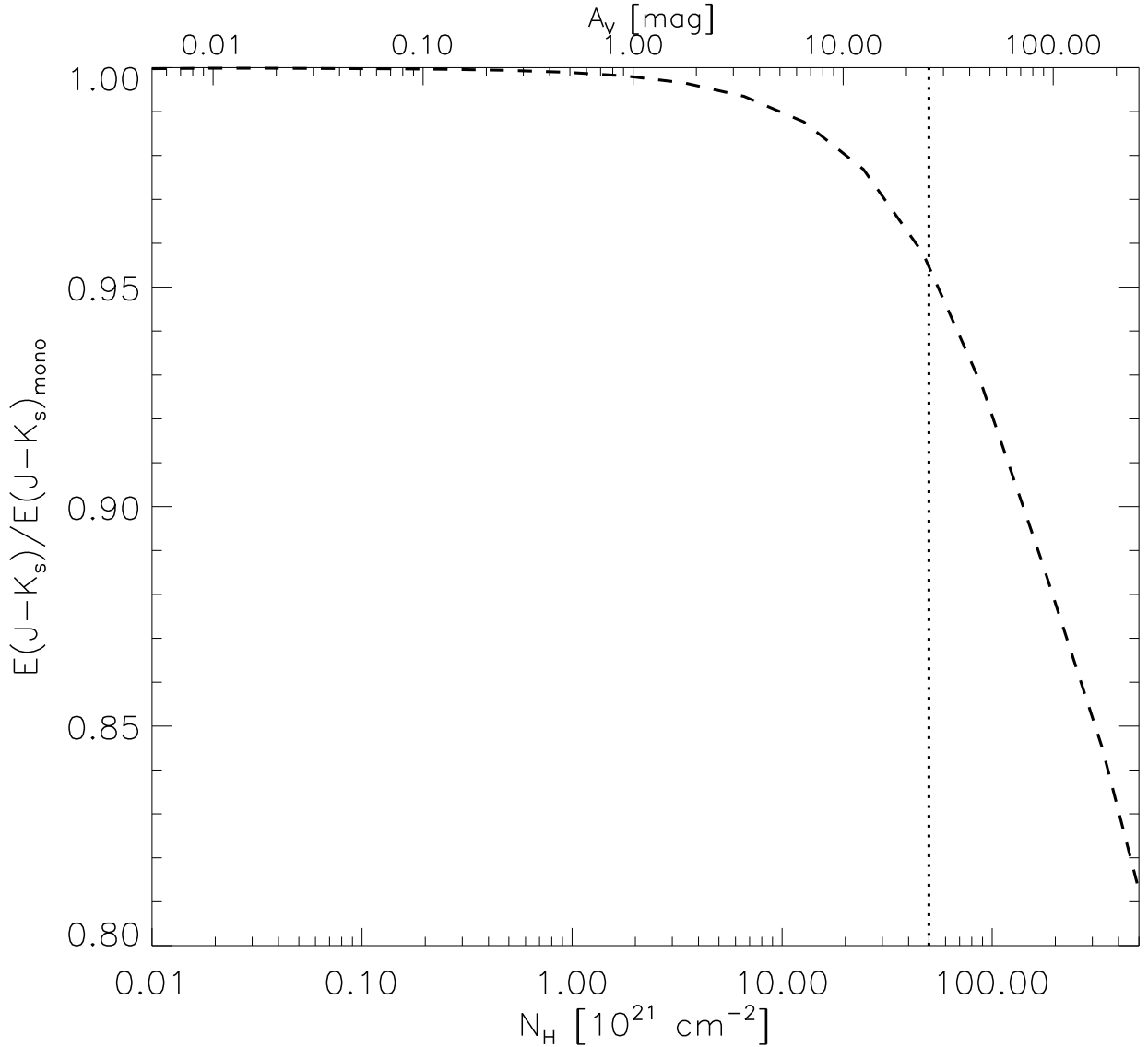}
\caption{Ratio of synthetic color-excess \Ejk\ obtained using 2MASS
  filters to the corresponding monochromatic color excess, as a
  function of \nh.  The vertical line shows the upper limit to \nh\ in
  the present analysis of lines of sight in the Orion~A molecular
  cloud, and so the effect on this study is small. }
\label{fig:color-correction}
\end{figure}

Specifically, we investigated the effect on the color
excess \Ejk\ which is the difference between extinction $A_{\rm J}$
and $A_{\rm K_s}$. 
Figure~\ref{fig:color-correction} shows the dependence of the ratio of
the apparent color excess \Ejk\ and the monochromatic color excess on
\nh, where \nh\ is obtained from the monochromatic color excess using
Equation~(\ref{eq:nhvsejk}).
Even for the lines of sight in Orion~A with the highest column density,
$N_{\rm H} \sim 50 \times 10^{21}$ cm$^{-2}$, the apparent color
excess (and the column density that would be deduced from it) is lower by
only 5\%.
For this calculation, we used $\alpha = 1.8$ and $\gamma = 1$, but the
results for these column densities are not sensitive to these choices.

\section{Predicted 100~\micron\ Emission} \label{appen:iris}

From the SED fits pixel by pixel to the four longest wavelength \her\ bands
(Section~\ref{sec:taut}), we have predicted the 100~\micron\ brightness
and then convolved and regridded this predicted map for comparison
with the 100~\micron\ IRIS image (Improved Reprocessing of the \IRAS\
Survey; \citealp{mairis}).\footnote{Alternatively, we convolved and
  regridded the SPIRE and PACS images to the IRIS resolution and then
  fit SEDs pixel by pixel at this coarser resolution, before
  predicting the 100~\micron\ emission; no systematic change was
  found.}
We compared the maps in two ways.  First, we computed the power
spectrum of each.  The shapes of the respective power spectra were 
very similar. Their relative amplitude indicated that  
the predicted map was 0.93 times fainter on average, very close 
considering that we made no color corrections for the finite \IRAS\ bandpass.

\begin{figure}
\includegraphics[width=\linewidth]{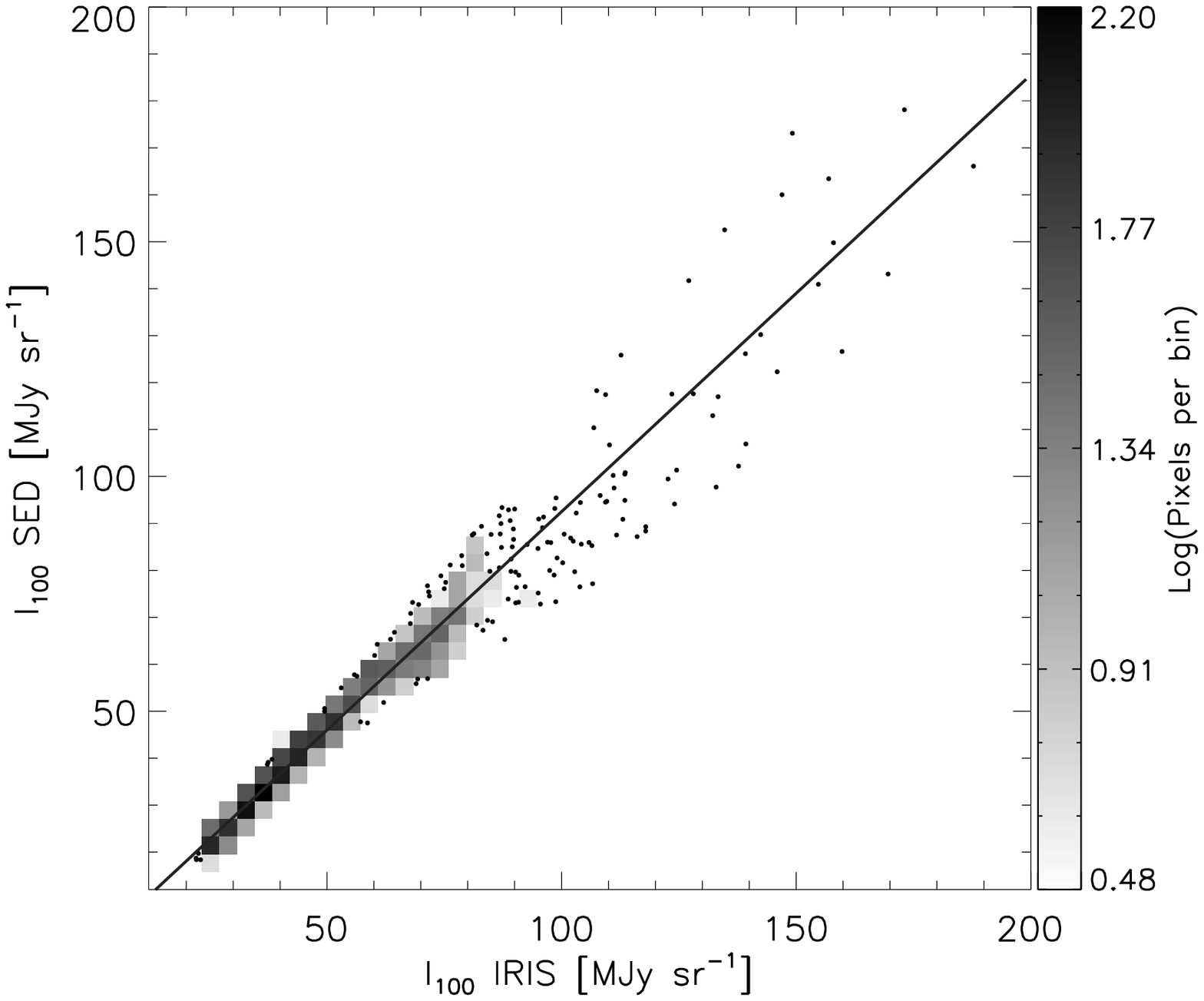}
\caption{ Correlation between the predicted 100~\micron\ intensity
obtained by extrapolating the fitted SED and the intensity directly
from the 100~\micron\ IRIS image.  The slope $0.93 \pm 0.01$ of the
correlation line is close to that anticipated from comparison of the
power spectra. Pixels representing  two-dimensional histogram is 
same as described in Figure \ref{fig:nhvsav}.
 Individual data points are plotted where the density of
points is less than or equal to three.}
\label{fig:iris}
\end{figure}
 
Second, we made a scatter plot for a pixel by pixel comparison of the
brightness in each map.  Figure~\ref{fig:iris} shows that the
correlation is very good.  For a formal linear fit we used the IDL
routine \emph{SIXLIN} \citep{isobe1990}, adopting the bisector.  The
slope $0.93 \pm 0.01$ is the same as obtained from the power spectra
and the intercept $-1.5\pm 0.5$~MJy~sr$^{-1}$ is close to zero as
expected.
The deviation of data from the correlation line at low surface
brightness  ($<$ 25 MJy sr$^{-1}$) is probably related to the
uncertainty associated with the zero-point offsets applied to the
\her\ maps (Appendix~\ref{sec:errors}).
This comparison validates the SED parameters obtained using only \her\
data and is consistent with the expectation that most of the emission
at 100~\micron\ is due to thermal emission from big grains.

We carried out a similar exercise for 70~\micron\ and 60~\micron\
emission and found that the brightness was significantly
underpredicted, as anticipated because of the contribution of
non-equilibrium emission by smaller grains.  The PACS datum plotted in
Figure~\ref{fig:sed} illustrates this point.


\section{Assessment of Uncertainties in the SED Parameters}\label{sec:errors}

Errors on the parameters from the SED fit were estimated by Monte
Carlo simulation \citep{chapin2008}.  For this assessment, we needed to 
estimate various sources of error in the intensities of the \her\ images.
If these errors are realistic, then a typical SED fit would have a
reduced $\chi^2$ close to unity and a map of $\chi^2$ would be
featureless.

We derived the minimum absolute error for the intensities in each
map from the power spectrum analysis as described in
Section~\ref{sec:obs}.  It was clear that a constant error across the
map of this order was too small in areas of high brightness; for the
latter, a constant percentage error of order 5\% was required to produce the
expected `flat' behavior in the $\chi^2$ map.

While investigating the details of the fits of individual SEDs for the
Orion~A-C1 field we found that the 350 and 500~\micron\ data points
were consistently above and below the best-fit SEDs for surface
brightnesses less than 25 and 10 MJy~sr$^{-1}$, respectively, and the
systematic deviation increased with decreasing surface
brightness. This motivated us to refine the offsets in the 350 and
500~\micron\ maps by correlating with the 250~\micron\ data.  For the
Orion~A-C1 field the correlation line between 250 and 160~\micron\
passed through the origin as expected.  The correlation lines
of 350 and 500~\micron\ relative to 250~\micron\ had intercepts of 3
and $-1.5$~MJy~sr$^{-1}$, respectively.  These values are of the magnitude
and sign that would produce the observed systematic deviations from
the best-fit SEDs for low surface brightness pixels.  The two
long-wavelength maps were therefore corrected by an additive offset,
making the intercepts formally zero.
A similar exercise on the Orion~A-S1 field also led us to correct the
350, 500, and as well as 160~\micron\ images by 2.5, 1, and
$-12.0$~MJy~sr$^{-1}$, respectively.  These refined offsets are not
perfect and we estimate an error approximately 10\% of the originally
applied zero-point offsets from \Planck\ and \IRAS.

Finally, we estimated the total error in the intensity to be fit by
adding in quadrature the minimum absolute error, 10\% of the offset value,
and 5\% of the intensity.  These estimates ought to be conservative
and do produce a flat $\chi^2$ image near two (as expected for four
intensities, two parameters, and thus two degrees of freedom).  Note
that the actual values of the derived parameters are not sensitive to
the precise details of the estimate of total noise.

In the Monte Carlo simulation of the uncertainties, 500 realizations
of mock data were generated by adding independent Gaussian noise
consistent with the above values to the actual intensities.  For each
realization, the SED was fit and the corresponding parameters were
recorded.  An uncertainty on each parameter was obtained by fitting a
Gaussian to the histogram of the generated distribution.  By keeping a
record of each fit, we also kept track of the correlations of the
uncertainties and so can produce the elliptical 1-$\sigma$ confidence
intervals in, for example, the $T$ -- $\tau_{1200}$ plane (see
Figure~\ref{fig:tvstau}).




\end{document}